\documentclass{article}
\usepackage[utf8]{inputenc}
\usepackage[a4paper, total={6in, 8in}]{geometry}
\usepackage{amsmath}
\usepackage{amssymb}
\usepackage{graphicx}
\usepackage{hyperref}
\usepackage{algorithm}
\usepackage[noend]{algpseudocode}
\usepackage{setspace}
\algrenewcommand\algorithmiccomment[1]{\hfill #1}
\usepackage{caption}
\usepackage{subcaption}
\usepackage{tikz}
\usepackage{tikzscale}
\usepackage{enumerate}
\usepackage[shortlabels]{enumitem}
\usepackage{boldline}
\usepackage{amsthm}
\usepackage{cleveref}
\usepackage{wrapfig}

\usepackage{authblk}

\newcommand{\R}{\mathbb{R}}
\newcommand{\E}{\mathbb{E}}
\newcommand{\ub}[1]{^{(#1)}}
\newcommand{\mat}[1]{\mathrm{#1}}
\newcommand{\id}{\mathbb{I}}
\newcommand{\T}{\mathrm{T}}

\newcommand{\gauss}{\mathbf{N}}

\newcommand{\sss}[1]{{\scriptscriptstyle #1}}

\DeclareMathOperator*{\argmax}{arg\,max}

\DeclareMathAlphabet{\mathcal}{OMS}{cmsy}{m}{n}

\newcommand*\return{\State \textbf{return} }

\author[1,2]{Samuel Duffield \footnote{Corresponding email: s@mduffield. Work completed whilst at the University of Cambridge.}}
\author[2]{Sumeetpal S. Singh}
\affil[1]{Quantinuum}
\affil[2]{University of Cambridge}

\title{\textbf{Quasi-Newton Sequential Monte Carlo}}
\date{November 2022}

\begin{document}

\maketitle

\begin{abstract}
    Sequential Monte Carlo samplers represent a compelling approach to posterior inference in Bayesian models, due to being parallelisable and providing an unbiased estimate of the posterior normalising constant. In this work, we significantly accelerate sequential Monte Carlo samplers by adopting the L-BFGS Hessian approximation which represents the state-of-the-art in full-batch optimisation techniques. The L-BFGS Hessian approximation has only linear complexity in the parameter dimension and requires no additional posterior or gradient evaluations.
    \par
    The resulting sequential Monte Carlo algorithm is adaptive, parallelisable and well-suited to high-dimensional and multi-modal settings, which we demonstrate in numerical experiments on challenging posterior distributions.
\end{abstract}

\section{Introduction}

Approximating complex probability distributions is a fundamental task in computational statistics, motivated in particular by inference in Bayesian models. In the Bayesian setting, we are tasked with approximating the posterior distribution
\begin{equation*}
    p(x \mid y) = \frac{p(x) p(y\mid x)}{p(y)},
\end{equation*}
where $y$ is a dataset, $p(y\mid x)$ is a \textit{likelihood} function mapping some unknown parameters $x$ to the known dataset $y$ and $p(x)$ is a prior distribution encoding pre-experimental knowledge about the parameter $x \in \R^d$.
In this work, we assume we have access to pointwise evaluations of the prior and likelihood but not of the normalising constant $p(y) = \int p(x) p(y\mid x) dx$. We also assume that we can access to gradients $\nabla \log p(x \mid y)$ either analytically or more likely through modern automatic differentiation techniques \cite{Margossian2019, jax}, note that by taking gradients in $\log$ space we need not evaluate the intractable normalising constant.
\par
The main goal of posterior approximation is to quantify expectations of the form $\E_{p(x \mid y)}[f(x)]$, such as predictions over test data.
\par
For complex or high-dimensional Bayesian models, quantifying the posterior represents a significant computational challenge. As such, a variety of methods for constructing a posterior approximation have been developed. Perhaps the simplest approach uses optimisation techniques to find a single \textit{maximum a posteriori} point estimate $x_\text{MAP} = \argmax_x p(x \mid y)$, this will be a poor approximation unless the posterior is very concentrated. An alternative approach is that of variational inference \cite{Blei2017} where a parameterised family of distributions is defined and then optimisation is used to find the optimal parameters based on some measure of distance between the posterior and variational distributions. This approach is computationally efficient, although induces a bias in the common case where the posterior lies outside of the variational family, furthermore this bias is difficult to quantify or control for general posteriors. Finally, there is the popular and flexible approach of developing a Monte Carlo approximation to the posterior $p(x \mid y) \approx \sum_{i=1}^N w\ub{i} \delta(x \mid x\ub{i})$ where $\delta$ is the Dirac point measure and $w\ub{i}$ are normalised weights $\sum_{i=1}^N w\ub{i} = 1$ (when $w\ub{i} = \frac1N \; \forall i$ the Monte Carlo approximation is said to be unweighted). Two dominant approaches for generating Monte Carlo approximations to Bayesian posteriors have developed, that of Markov chain Monte Carlo (MCMC) (e.g. \cite{Brooks2011}) and importance sampling (e.g. \cite{Chopin2020}). In MCMC, a sample is evolved according to a Markov kernel $q(x_t \mid x_{t-1})$ that is invariant for the posterior distribution, collecting samples along the Markov chain then provides an asymptotically unbiased and unweighted approximation to the posterior distribution. In contrast, importance sampling in its most simple form generates $N$ samples independently from a known proposal distribution $x\ub{i} \sim q(x)$ before correcting for the discrepancy between $q(x)$ and $p(x \mid y)$ with importance weights $w\ub{i} \propto \frac{p(x\ub{i} \mid y)}{q(x\ub{i})}$, again this provides an asymptotically unbiased Monte Carlo approximation. In practice, it is difficult to construct efficient importance sampling proposal distributions and as a result, the importance weights are often dominated by a single sample, this is in contrast to MCMC proposals which can leverage gradients to construct efficient exploration of the posterior.
\par
Sequential Monte Carlo (SMC) \cite{Chopin2020} represents an extension of the importance sampling paradigm that introduces intermediate distributions to gradually approach the posterior and maintain stability in the importance weights. SMC was first introduced for the specific case of state-space models \cite{Gordon93} but then later generalised to static Bayesian posteriors \cite{DelMoral2006}. By adopting this sequential approach, SMC can utilise efficient gradient-based proposal distributions from MCMC in order to explore the intermediate distributions. Notably, in contrast to MCMC, importance sampling and SMC approaches are parallelisable (making them well suited to modern GPU/TPU architectures) and provide an approximation to the posterior normalisation constant which is particularly useful for Bayesian model selection \cite{Hoeting1999}.
\par
In this work, we investigate the use of a Hessian approximation to accelerate sequential Monte Carlo for static Bayesian posterior inference. The Hessian matrix represents a natural pre-conditioner, capturing the local scaling and correlations of the target distribution, however calculating the Hessian matrix exactly is computationally expensive and not guaranteed to be symmetric positive definite, which is typically a requirement for pre-conditioned sampling. We instead adopt a quasi-Newton method (L-BFGS), ubiquitous in optimisation literature, to convert gradient evaluations from a particle's historical trajectory into a computationally cheap, positive definite projection of the Hessian matrix.
\par
Sequential Monte Carlo for static Bayesian inference with tempered intermediate distributions is reviewed in \Cref{sec:lik_temp}. In \Cref{sec:langevin} we detail how to incorporate gradient information into a Markovian transition kernel. We then describe the L-BFGS Hessian approximation and its implementation within the transition kernel in \Cref{sec:qn_langevin}. \Cref{sec:sims} presents two challenging numerical experiments, a high-dimensional toy example with difficult scaling and a multi-modal target distribution with real data. \Cref{sec:discu} provides discussion and extensions.

\section{Likelihood Tempering}\label{sec:lik_temp}
Sequential Monte Carlo requires a sequence of target distributions $\pi_\sss{0:t}(x_{0:t}),$ $t=0, \dots, T$. For static Bayesian inference, the sequence is typically defined artificially such that the marginal distribution of particles at the final iteration is the posterior distribution $\pi_T(x) = p(x\mid y) \propto p(x)p(y \mid x)$.
\par
In this work we will consider the case of \textit{likelihood tempering} \cite{Gelman1998, Neal2001} where we fix the intermediate marginal distributions
\begin{align*}
    \pi_t(x) &\propto p(x)p(y \mid x)^{\lambda_t} \propto \exp(-U_t(x)),
\end{align*}
where $U_t(x) = -\log p(x) - \lambda_t \log p(x \mid y)$ is the \textit{tempered potential} and $\lambda_t$ is an \textit{inverse temperature} parameter increasing at each iteration
\begin{equation*}
    0 \leq \lambda_0 < \lambda_1 < \dots < \lambda_T = 1.
\end{equation*}
Under this construction, the sequence `smoothly' transitions from the prior to the posterior.
\par
Alternative intermediate targets are possible. For example, in the case where we receive $T$ observations $(y_1, \dots, y_T)$ we can set batched intermediate targets $\pi_t(x_t) = p(x_t \mid y_1, \dots, y_t)$ as in \cite{Chopin2002}. However, this strategy does not necessarily induce smooth transitions between the intermediate target distributions.

\par
Although not investigated here it would be possible to consider alternative stopping criteria that extend to $\lambda_T >1$ with the goal of optimisation rather than integration \cite{Duffield2022}.
\par

\subsection{Sequential Importance Weights}

Having fixed the marginal targets $\pi_t(x_t)$, we can assume we have weighted particles at time $t-1$ approximating $\pi_{t-1}(x_{t-1})$. Suppose we use a Markovian transition kernel
\begin{align*}
    q_\sss{t|0:t-1}(x_t \mid x_{0:t-1}) = q_{t|t-1}(x_t \mid x_{t-1}),
\end{align*}
then we can write the sequential weights as
\begin{align*}
    w_t &= \frac{
    \pi_t(x_t) \pi_{t-1|t}(x_{t-1} \mid x_t)
    }{
    q_{t-1}(x_{t-1}) q_{t|t-1}(x_t \mid x_{t-1})
    }, \\
    &= \frac{\pi_{t-1}(x_{t-1})}{q_{t-1}(x_{t-1})}
    \frac{
    \pi_t(x_t) \pi_{t-1|t}(x_{t-1} \mid x_t)
    }{
    \pi_{t-1}(x_{t-1}) q_{t|t-1}(x_t \mid x_{t-1})
    }, \\
    &= w_{t-1} \frac{
    \pi_t(x_t) \pi_{t-1|t}(x_{t-1} \mid x_t)
    }{
    \pi_{t-1}(x_{t-1}) q_{t|t-1}(x_t \mid x_{t-1})
    }.
\end{align*}
where $\pi_{t-1|t}(x_{t-1} \mid x_t)$ is a normalised backward kernel which we are free to choose and $q_{t-1}(x_{t-1})$ is the intractable, exact marginal distribution of $x_{t-1}$.
\par
Further suppose that the Markovian transition kernel is $\pi_{t-1}$-invariant 
\begin{align*}
    \int q_{t|t-1}(x_t \mid x_{t-1}) \pi_{t-1}(x_{t-1}) dx_{t-1} = \pi_{t-1}(x_{t}).
\end{align*}
Then the natural choice of $\pi_{t-1|t}(x_{t-1} \mid x_t)$, \cite{DelMoral2006}, is
\begin{align*}
    \pi_{t-1|t}(x_{t-1}|x_t) =
    \frac{\pi_{t-1}(x_{t-1}) q_{t|t-1}(x_t | x_{t-1})}
    {\pi_{t-1}(x_{t})},
\end{align*}
which induces sequential importance weights of the form
\begin{align*}
    w_t &= w_{t-1}\frac{\pi_t(x_t)}{\pi_{t-1}(x_t)}.
\end{align*}
This procedure is detailed (in self-normalising form) in steps 16-17 of Algorithm~\ref{alg:met_smc}.


\subsection{Adaptive Tempering}\label{adapt_temp}
Combining the sequential importance weights induced from a $\pi_{t-1}$-invariant, Markovian transition kernel with likelihood tempered intermediate distributions gives weights
\begin{equation}\label{temper_weights}
    w_t = w_{t-1}p(y\mid x_t)^{\lambda_t - \lambda_{t-1}}.
\end{equation}
A major advantage of this formulation is that the weights $w_t$ can be evaluated as a function of $\lambda_t$ without any further likelihood evaluations, \cite{Jasra2011}. Thus, we can define the inverse temperature schedule adaptively by using a numerical root finder (i.e. bisection) at each iteration to solve
\begin{equation*}
    \text{ESS}(\lambda_t) \approx \rho \text{ESS}(\lambda_{t-1}).
\end{equation*}
For $\lambda_t \in (\lambda_{t-1}, 1]$ where $\rho$ is a design parameter controlling the $\chi^2$-distance between  $\pi_{t-1}$ and $\pi_{t}$ \cite{Chopin2020} and therefore the number of iterations, $T$, required to reach the posterior $\pi_T(x) = p(x \mid y)$. This is step 15 in Algorithm~\ref{alg:met_smc}. We define the effective sample size as
\begin{equation*}
     1 \leq \text{ESS}(\lambda_t) = \frac{\left(\sum_{i=1}^N w_t\ub{i}(\lambda_t) \right)^2}{\sum_{i=1}^N w_t\ub{i}(\lambda_t)^2} \leq N,
\end{equation*}
which is standard in SMC although other sample quality metrics are applicable \cite{Elvira2022}. Here $w_t\ub{i}(\lambda_t)$ represents the weight in \eqref{temper_weights}, for the $i$th particle, as a function of the next inverse temperature $\lambda_t$.

\subsection{Resampling}
In order to maintain a diverse collection of particles, SMC samplers \cite{DelMoral2006} apply a \textit{resampling} operation when the effective sample size becomes small, i.e. when $\text{ESS}(\lambda_{t}) < \kappa N$ for a second threshold parameter $\kappa$.
\par
The resampling operation rejuvenates the particles whilst maintaining the asymptotic unbiasedness of the approximation. It can be applied in many ways \cite{Douc2005}. In this work, we adopt the simplest multinomial resampling where particles are sampled with replacement from the population according to their weights, the post-resampling weights are then set to the uniform $1/N$. This is detailed in steps 8-11 of Algorithm~\ref{alg:met_smc}.

\section{Langevin Kernel}\label{sec:langevin}
We now turn to the choice of transition kernel. Assuming we have access to gradients $\nabla U(x)$ we can adopt a gradient-informed proposal based on the overdamped Langevin diffusion
\begin{equation*}
    dx_t = - \nabla U(x_t) dt + \sqrt{2}dW_t,
\end{equation*}
where $W_t$ is a standard Brownian motion. The continuous-time dynamics are invariant for $\pi(x) \propto \exp(-U(x))$ \cite{Ma2015}, however this property is not retained for the Euler-Maruyama discretisation
\begin{equation*}
    q(x \mid x_{t-1}) = \gauss (x \mid x_{t-1} - \epsilon \nabla U(x_{t-1}), 2 \epsilon \id_d). 
\end{equation*}
Invariance can be regained with the use of a Metropolis-Hastings step where a proposal $x' \sim q(x \mid x_{t-1})$ is accepted $x_t = x'$ with probability
\begin{equation}\label{mh}
    \alpha(x_{t-1}, x') = \min\left(1,
    \frac{\pi(x') q(x_{t-1} \mid x')}{\pi(x_{t-1})q(x' \mid x_{t-1})}
    \right),
\end{equation}
otherwise, the proposed $x'$ is rejected and the previous particle is duplicated $x_t = x_{t-1}$. The resulting MCMC algorithm is referred to as the \textit{Metropolis Adjusted Langevin Algorithm} (MALA) \cite{Roberts1996}.
\par
The only tuning parameter of this kernel is the stepsize $\epsilon$. Sophisticated adaptive schemes for tuning the stepsize within sequential Monte Carlo have been developed in \cite{Buchholz2021}, in this work we use a Robbins-Monro algorithm (with constant adaptation stepsize $\delta$) to ensure the average Metropolis-Hastings acceptance probability $\bar{\alpha}_t = \frac1N \sum \alpha_t\ub{i}$ is pushed towards a target $\alpha^*$. The full sequential Monte Carlo regime including stepsize adaptation and generic proposal distribution is detailed in \Cref{alg:met_smc}.

\begin{algorithm}
\caption{Metropolised SMC}\label{alg:met_smc}
\begin{algorithmic}[1]\onehalfspacing
\State Sample from prior $x_0\ub{i} \sim p(x_0)$ \Comment{$i=1\dots,N$}
\State Solve for $\lambda_0 \in (0, 1]$ such that $\text{ESS}(\lambda_0) \approx \rho N$
\State Normalise $\hat{Z}_0 = \frac1N \sum_{i=1}^N p(y \mid x_0\ub{i})^{\lambda_0}$
\State Weight $w_0\ub{i} = \frac{p(y \mid x_0\ub{i})^{\lambda_0}}{N\hat{Z}_0}$\Comment{$i=1\dots,N$}
\State Set $t=0$
\While{$\lambda_t\leq 1$}
\State Set $t=t+1$
\If{$\text{ESS}(\lambda_{t-1}) < \kappa N$}
\State $\left\{ \tilde{x}_{0:t-1}\ub{i}, \tilde{w}_{t-1}\ub{i}=\frac1N \right\}_{i=1}^N = \textrm{Resample}\left( \left\{x_{0:t-1}\ub{i}, w_{t-1}\ub{i}\right\}_{i=1}^N\right)$
\Else
\State $\left\{\tilde{x}_{0:t-1}\ub{i}, \tilde{w}_{t-1}\ub{i} \right\}_{i=1}^N = \left\{x_{0:t-1}\ub{i}, w_{t-1}\ub{i} \right\}_{i=1}^N$
\EndIf
\For{$i=1,\dots,N$}
\State Sample $\tilde{x}_t\ub{i} \sim q_{t-1}(x \mid \tilde{x}_{t-1}\ub{i})$
\State With probability $\min \left(1,
\frac{\pi_{t-1}(\tilde{x}_t\ub{i})q_{t-1}(\tilde{x}_{t-1}\ub{i} \mid \tilde{x}_{t}\ub{i})
}{
    \pi_{t-1}(\tilde{x}_{t-1}\ub{i})q_{t-1}(\tilde{x}_{t}\ub{i} \mid \tilde{x}_{t-1}\ub{i})
    } \right)$
    set $x_t\ub{i} = \tilde{x}_{t}\ub{i}$,
\Statex \hspace{7.64cm} otherwise $x_t\ub{i} = \tilde{x}_{t-1}\ub{i}$.
\EndFor
\State Solve for $\lambda_t \in (\lambda_{t-1}, 1]$ such that $\text{ESS}(\lambda_t) \approx \rho \text{ESS}(\lambda_{t-1})$
\State Normalise $
\hat{Z}_\sss{t|0:t-1}= \sum_{i=1}^N \tilde{w}_{t-1}\ub{i} p(y \mid x_t\ub{i})^{\lambda_t}
$
\State Reweight
\begin{equation*}
w_t\ub{i} = 
\tilde{w}_{t-1}\ub{i}
\frac{
p(y \mid x_t\ub{i})^{\lambda_t}
}{
\hat{Z}_\sss{t|0:t-1}}
\tag*{$i=1\dots,N$}
\end{equation*}
\State Adapt stepsize $\log \epsilon_{t+1} = \log \epsilon_t + \delta (\bar{\alpha}_t - \alpha^*)$
\EndWhile
\return $\left\{ \left\{x\ub{i}_{t}, w_t\ub{i}\right\}_{i=1}^N \right\}_{t=0}^T$
\end{algorithmic}
\end{algorithm}

\section{Quasi-Newton Langevin Kernel}\label{sec:qn_langevin}
The idea behind \cite{Girolami2011} is to extend gradient-based Markov Chain Monte Carlo methods to the case of Langevin dynamics with a position-dependent preconditioner. In the overdamped case, we get
\begin{equation*}
    dx_t = - \Sigma(x_t)\nabla U(x_t) dt + \text{div}(\Sigma(x_t))dt + \sqrt{2\Sigma(x_t)}dW_t,
\end{equation*}
for position-dependent preconditioner $\Sigma(x_t)$ and matrix derivative $[\text{div}(\Sigma(x))]_i = \sum_{j=1}^d \frac{\partial}{\partial x_j} \Sigma_{ij}(x)$. This continuous-time process is invariant for $\pi(x) \propto \exp(-U(x))$ \cite{Ma2015}.
\par
In practice, the matrix derivative term $\text{div}(\Sigma(x_t))$ is typically intractable or too expensive to calculate, as such we consider \textit{simplified pre-conditioned overdamped Langevin dynamics}
\begin{equation*}
    dx_t = - \Sigma(x_t)\nabla U(x_t) dt + \sqrt{2\Sigma(x_t)}dW_t,
\end{equation*}
with the matrix terms omitted. Invariance can still be regained through the use of a Metropolis-Hastings step \eqref{mh} on an Euler-discretised proposal
\begin{equation}\label{precon_prop}
    q(x \mid x_{t-1}) = \gauss (x \mid x_{t-1} - \epsilon \Sigma(x_{t-1})\nabla U(x_{t-1}), 2 \epsilon \Sigma(x_{t-1})). 
\end{equation}
\par
A logical choice for the preconditioner $\Sigma(x)$ is the inverse Hessian matrix $\Sigma(x) = (\nabla^2 U(x))^{-1}$. This choice is logical as it induces affine invariant dynamics \cite{Leimkuhler2018}, that is the performance of the sampler is unchanged under a linear reparameterisation.
\par
Unfortunately, we cannot easily use the Hessian matrix directly for general target distributions. Firstly, the Hessian matrix is not necessarily symmetric, positive-definite or invertible and secondly the required matrix inversion and square root operations comes at a prohibitive cost of $O(d^3)$ for $x\in \R^d$.

\subsection{L-BFGS}

Ref. \cite{Zhang2011} suggest overcoming these issues by invoking a Hessian approximation based on the Broyden–Fletcher–Goldfarb–Shanno (BFGS) algorithm \cite{Nocedal2006}. The BFGS algorithm and its limited-memory variant (L-BFGS) represent the state-of-art in non-stochastic optimisation where a sequence of iterates $x_t = x_{t-1} - \epsilon_t \mat{B}^{-1}_{t-1} \nabla U(x_{t-1})$ is generated to minimise the function $U(x)$ where $\mat{B}_{t-1} \approx \nabla^2 U(x_{t-1})$ is an approximation to the Hessian at $x_{t-1}$ - thus representing a so called \textit{quasi-Newton} method.
\par
We utilise the following L-BFGS recursion \cite{Zhang2011} which directly approximates the square-root of the Hessian and its inverse
\begin{subequations}\label{bfgs}
\begin{align}
    \mat{B}_{t+1} &= \mat{C}_{t+1} \mat{C}_{t+1}^\T, &
    \mat{B}^{-1}_{t+1} &= \mat{S}_{t+1} \mat{S}_{t+1}^\T,\\
    \mat{C}_{t+1} &= (\id_d - \mathrm{u}_t \mathrm{t}_t^\T) \mat{C}_{t}, &
    \mat{S}_{t+1} &= (\id_d - \mathrm{p}_t \mathrm{q}_t^\T) \mat{S}_{t}, \\
    \mathrm{t}_t &= \frac{s_t}{s_t^\T \mat{B}_t s_t}, &
    \mathrm{p}_t &= \frac{s_t}{s_t^\T y_t},  \\
    \mathrm{u}_t &= \sqrt{\frac{s_t^\T \mat{B}_t s_t}{s_t^\T y_t}} y_t + \mat{B}_t s_t &
    \mathrm{q}_t &= \sqrt{\frac{s_t^\T y_t}{s_t^\T \mat{B}_t s_t}} \mat{B}_t s_t + y_t,
\end{align}
\end{subequations}
where $s_t = x_{t+1} - x_{t}$ and $y_t = \nabla U(x_{t+1}) - \nabla U(x_{t})$. The limited-memory variant only applies $m$ steps of the recursion initiated with diagonal matrices $\mat{C}_{t-m} = \mat{S}_{t-m}^{-1}$ that represent an initial guess for the square root of the Hessian and inverse.
\par
In practice, the vectors $\left\{\mathrm{p}_r, \mathrm{q}_r, \mathrm{u}_r, \mathrm{t}_r \right\}_{r=t-m}^t$ are pre-computed at a cost of $O(m^2d)$ and subsequent matrix vector products can be computed at a cost of $O(md)$ using the sequence of inner products
\begin{align*}
    \mat{C}_{t+1}z
    &= (\id_d - \mathrm{u}_t \mathrm{t}_t^\T) \dots (\id_d - \mathrm{u}_{t-m+1} \mathrm{t}_{t-m+1}^\T) \mat{C}_{t-m}z,\\
    \mat{S}_{t+1}z
    &= (\id_d - \mathrm{p}_t \mathrm{q}_t^\T) \dots (\id_d - \mathrm{p}_{t-m+1} \mathrm{q}_{t-m+1}^\T) \mat{S}_{t-m}z,
\end{align*}
thus the L-BFGS variant as described above provides access to a factorised approximation of the Hessian and inverse Hessian matrices all at a cost that is \textbf{linear} in dimension.
\par
We are yet to ensure the Hessian approximation is invertible (or rather positive definite and therefore invertible). We can do this by checking that $s_r^\T y_r >0$ for each $r=t-m, \dots, t$ \cite{Nocedal2006}. This is obtained in \cite{Zhang2011} by simply removing points from the recursion when $s_r^\T y_r \leq0$. In this work, we adopt a strategy similar to \cite{Schraudolph2007} where we notice that we can instead approximate $\nabla^2 U(x) + \beta \mat{B}_{t-m}$ by adjusting each $y_r \leftarrow y_r + \beta\mat{B}_{t-m}s_r$. We can guarantee a positive definite approximation given $\beta$ suitably large and a positive definite (diagonal) initial guess $\mat{B}_{t-m}$. In practice, we adaptively set
\begin{equation*}
    \beta = \max\left(0, \max\left( \left\{ \frac{-s_r^\T y_r}{s_r^\T \mat{B}_{t-m} s_r}\right\}_{r=t-m}^t \right) + \omega\right),
\end{equation*}
so that each $s_r^\T y_r > \omega$ for some bounding parameter $\omega > 0$.
\par
We now have the tools we need to apply the preconditioned Langevin proposal \eqref{precon_prop} with preconditioner $\Sigma(x) = \mat{B}_t(x_{t-m-1:t-1})^{-1}$ where $\mat{B}_t(x_{t-m-1:t-1})$ represents the L-BFGS Hessian approximation at inverse temperature $\lambda_t$ using trajectory values $x_{t-m-1:t-1}$ - which is positive-definite and is accessible in inverse and factorised form.
\par
The use of the previous trajectory means the proposal is no longer Markovian. Ref. \cite{Zhang2011} correct for this to obtain a valid Markov Chain Monte Carlo sampler by extending the state to include all $m$ previous values - however, this approach is complicated when applied within a tempered SMC sampler. In this work, we do not correct for this bias and instead note as in \cite{Wang2021} that the bias is controllable as increasing $m$ and decreasing the stepsize $\epsilon$ increases the accuracy of the Hessian approximation (or rather its positive definite projection).

\section{Numerical Experiments}\label{sec:sims}

We now investigate the numerical performance of the sequential Monte Carlo regimes with both classical Langevin kernel and quasi-Newton Langevin kernel.
\par
It is common for sequential Monte Carlo to take multiple MCMC steps at each iteration \cite{Dau2020} alongside an aggressive choice of the ESS threshold parameter $\rho$ (i.e. $\rho=0.5$) that controls the size of temperature jumps. With an aggressive choice of $\rho$ many likelihood evaluations are taken at each iteration but the $\chi^2$-distance between tempered distributions is large. In this work, we take an alternative approach with only one MCMC step alongside modest $\rho=0.95$, this way only one likelihood evaluation is executed per particle per iteration. Although not utilised here, this approach has the advantage that every likelihood evaluation can contribute to expectation approximations via the waste-recycling technique used in \cite{Nguyen2016, Gramacy2010}. This technique uses the fact that the output from each iteration $\{x_t\ub{i}, w_t\ub{i}\}_{i=1}^N$ is asymptotically unbiased for the tempered intermediate distribution $\pi_t(x)$, and therefore we can adjust the weights to target the posterior $\pi_T(x)$, this way all $NT$ particles contribute to posterior expectations.
\par
In all experiments, we fix $N=1000$ and resample when the effective sample size falls below $0.5N$. As mentioned we use a Robbins-Monro schedule with constant adaptation stepsize $\delta=1$ to keep the Metropolis acceptance probability close to 80\%. In the L-BFGS subroutine, we fix the positive-definite parameter $\omega=1$ and use a memory size $m=20$. All experiments are repeated 20 times. Code is written in mocat \cite{mocat2021}.

\subsection{High Dimensional Gaussian}
The first inference task we consider exhibits high dimensionality and inhomogeneous scaling \cite{Neal2011} where the target distribution $\pi(x)$ is a 100-dimensional, zero-mean Gaussian with covariance $\mat{Q} = \text{diag}(0.01^2, \dots, 0.99^2, 1)$.
\par
As the problem does not have a prior-likelihood structure we use the following artificial likelihood tempering
\begin{equation*}
    \pi_t(x) = \pi_0(x) \left( \frac{\pi(x)}{\pi_0(x)} \right)^{\lambda_t},
\end{equation*}
with artificial prior $\pi_0(x) = \gauss(x \mid 0, \id_{100})$.
\par
In our L-BFGS implementation, we initiate the Hessian approximations with the inverse of the diagonal sample covariance of the previous particles.
\par

\begin{figure}
\begin{minipage}{0.48\linewidth}
    \centering
    \includegraphics[width=0.9\linewidth]{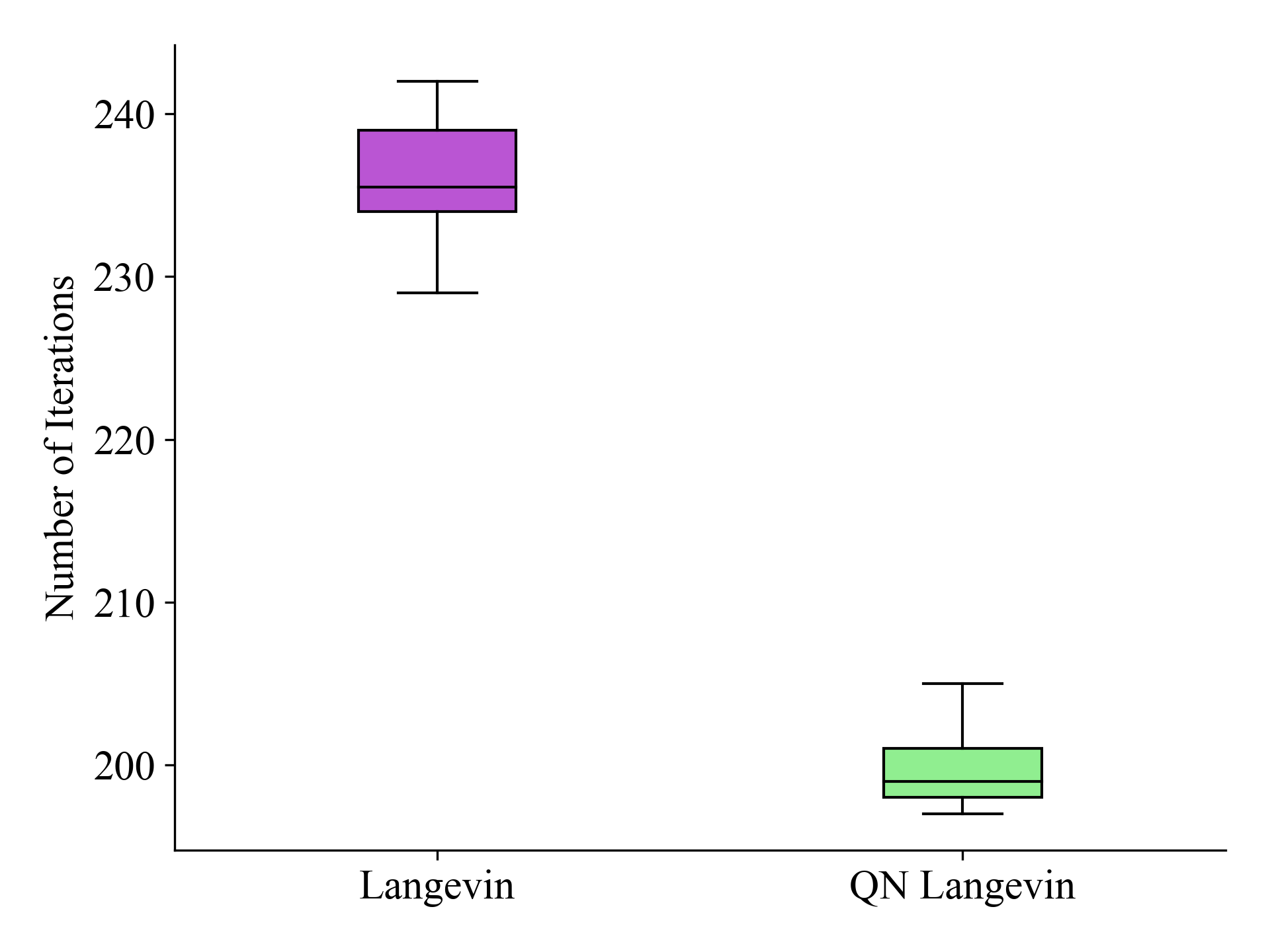}
    \caption{Number of iterations required to reach posterior in high dimensional Gaussian example.}
    \label{fig:gauss_temps}
\end{minipage}
\hfill
\begin{minipage}{0.48\linewidth}
    \centering
    \includegraphics[width=0.9\linewidth]{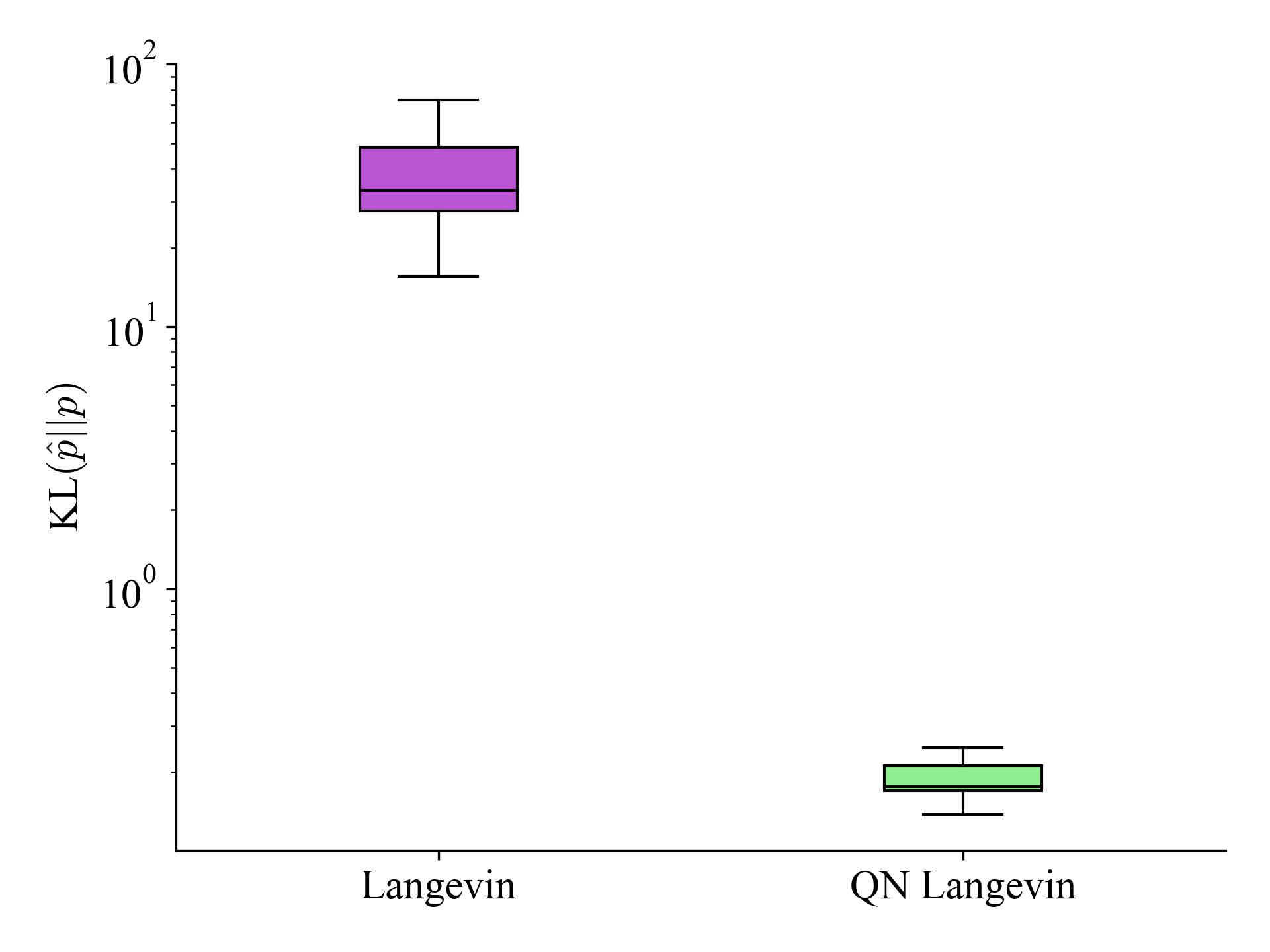}
    \caption{KL-divergence between final particle approximation in high dimensional Gaussian example.}
    \label{fig:gauss_kl}
\end{minipage}
\end{figure}

In \Cref{fig:gauss_temps}, we display the number of iterations, $T$, the adaptive sequential Monte Carlo schemes required to reach the target distribution at inverse temperature $\lambda_T=1$. The number of iterations required represents the difference in computational cost between the Metropolised classical proposal and the Metropolised quasi-Newton Langevin proposal - as both require the same number of likelihood and gradient evaluations per iteration. In \Cref{fig:gauss_kl}, we analyse the accuracy of the particle approximations by comparing the KL-divergence from the Gaussian distribution induced by the weighted sample mean and covariance of the final particles to the true target distribution $\gauss(x \mid 0, \mat{Q})$.
\par
We observe that the preconditioned sampler is faster in \Cref{fig:gauss_temps}, requiring fewer iterations and therefore fewer likelihood evaluations and resampling operations. We also see that the quasi-Newton proposal is substantially more accurate in \Cref{fig:gauss_kl} - note that the KL-divergence is displayed on a $\log$-scale. The quasi-Newton kernel has captured the difficult scaling whilst the classical Langevin has struggled to move away from homogeneity.

\subsection{Gaussian Mixture Model}

\begin{wrapfigure}[10]{r}{0.4\textwidth}
    \centering
    \includegraphics[width=0.9\linewidth]{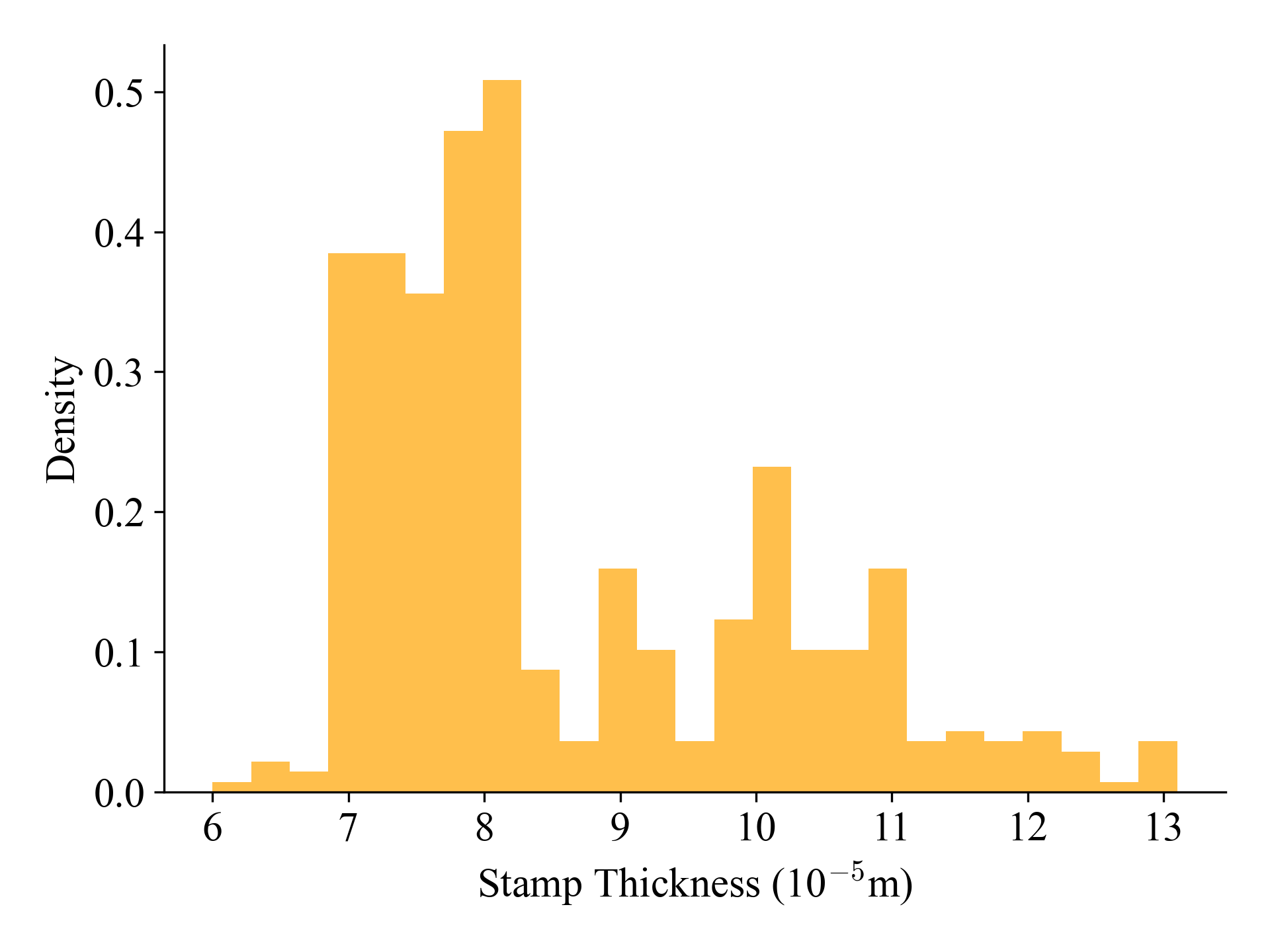}
    \caption{Hidalgo stamp data.}
    \label{fig:stamps}
\end{wrapfigure}

Our second example is considered in \cite{Chopin2012} and represents fitting a univariate dataset  $y = (y_1, \dots, y_K)$ with a weighted sum of three Gaussian distributions
\begin{equation*}
    p(y_k \mid z, \mu, \nu) = \sum_{i=1}^3 z_i \, \gauss (y_k \mid \mu_i, \nu_k^{-1}),
\end{equation*}
where $\sum_{i=1}^3 z_i = 1$. We define the prior distribution hierarchically
\begin{align*}
    (z_1, z_2, z_3) &\sim \text{Dirichlet}(1, 1, 1), \\
    \mu_i &\sim \gauss( \cdot \mid a, b^{-1}), \tag*{$i=1,2,3$,}\\
    \nu_i &\sim \text{Gamma}( \cdot \mid \alpha, \beta), \tag*{$i=1,2,3$,}\\
    \beta &\sim \text{Gamma}( \cdot \mid g, h),
\end{align*}
where the constants $a$, $b$, $\alpha$, $g$ and $h$ are set as it \cite{Leimkuhler2018}. Our parameter to be inferred is therefore the nine-dimensional $x = (\mu, \nu, z_1, z_2, \beta)$ with $z_3 = 1-z_1-z_2$.
\par
Note that we have the constraints $\nu_i > 0$, $\beta>0$ and a 3-simplex constraint on the weights $z$. To facilitate the gradient-based sampling algorithms we take $\log$ transforms on the positive parameters $\nu, \beta$ and use the simplex transformation detailed in \cite{Betancourt2012} to unconstrain the weights $z$. We adjust the prior density with the transformation Jacobians accordingly.

We consider fitting the Hidalgo stamp dataset \cite{Izenman1988} which consists of 485 data points representing the thickness of individual stamps, depicted in \Cref{fig:stamps}. This model exhibits a \textit{label switching problem} where a priori each of the three mixture components are identical and are therefore invariant to re-labelling. Thus the model admits $3!=6$ local modes which will each have their own local scaling.
\par
When calculating the L-BFGS Hessian approximations \eqref{bfgs}, we initiated the Hessian with the identity (due to the anticipated multi-modal behaviour).
\par

\begin{figure}
\begin{minipage}{0.48\textwidth}
    \centering
    \includegraphics[width=0.9\textwidth]{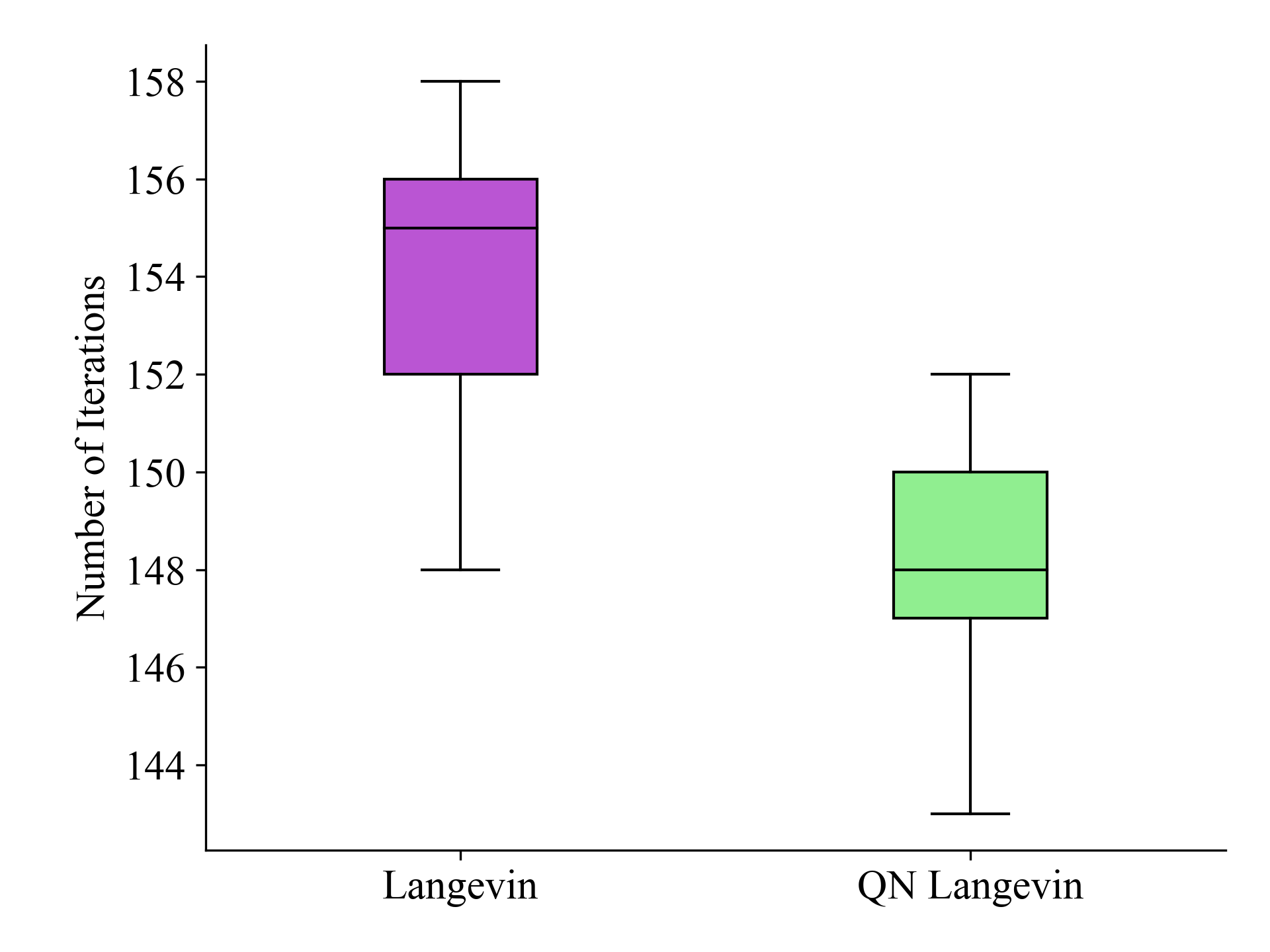}
    \caption{Number of iterations required to reach posterior in Gaussian mixture model example.}
    \label{fig:gmm_temps}
\end{minipage}
\hfill
\begin{minipage}{0.48\textwidth}
    \centering
    \includegraphics[width=0.9\textwidth]{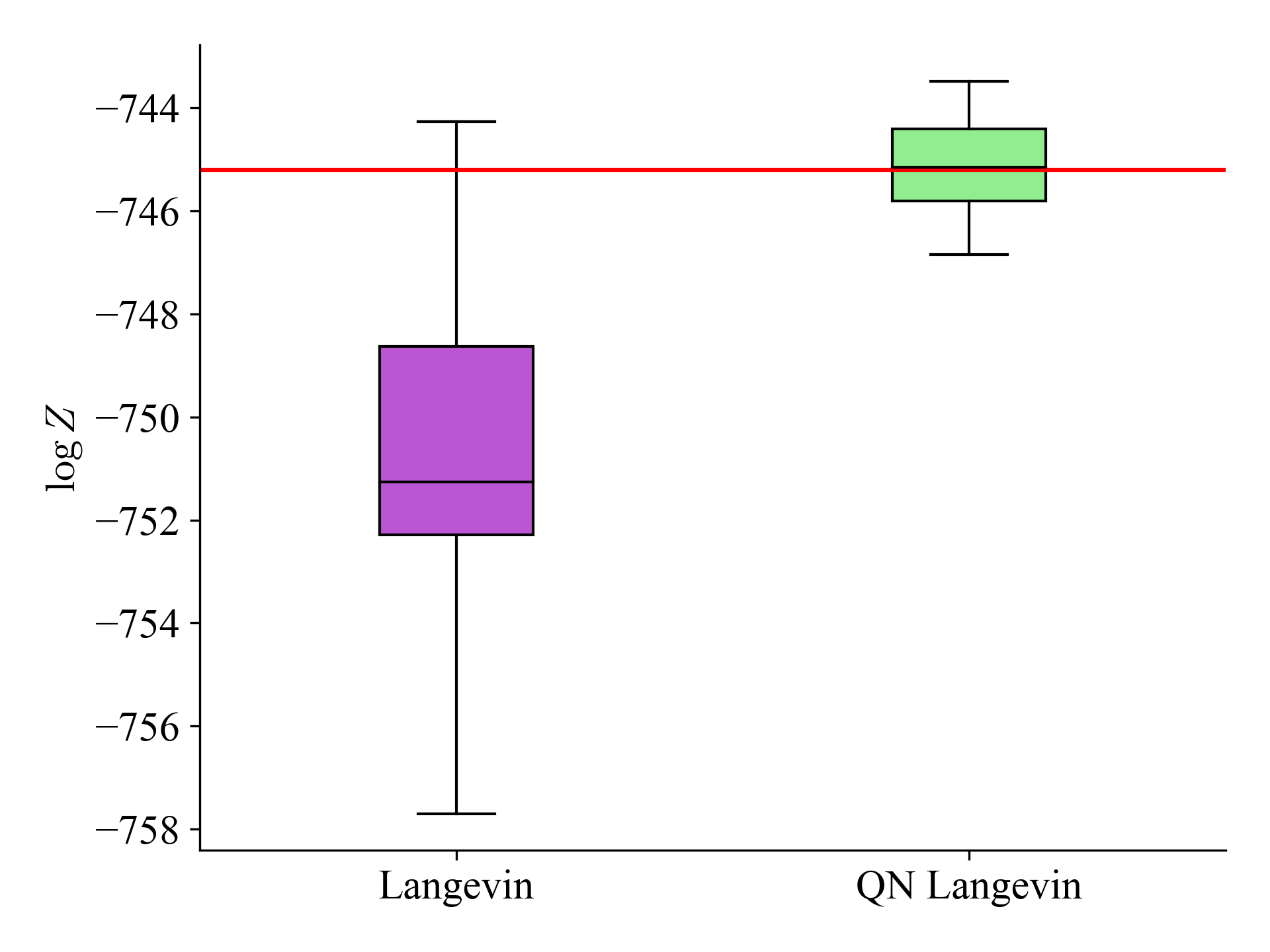}
    \caption{Estimate of log normalising constant in Gaussian mixture model example. Truth represented by horizontal red line.}
    \label{fig:gmm_logz}
\end{minipage}
\end{figure}

\Cref{fig:gmm_temps} displays the number of temperatures or iterations required to reach the posterior distribution - which is determined adaptively.
We again notice, as in the Gaussian example, that the quasi-Newton method accelerates more quickly to the posterior distribution resulting in fewer iterations (and therefore likelihood, gradient and resampling operations) than its classical counterpart.
\par
\begin{figure}
	\begin{minipage}{0.48\textwidth}
        \centering
        \includegraphics[width=0.9\textwidth]{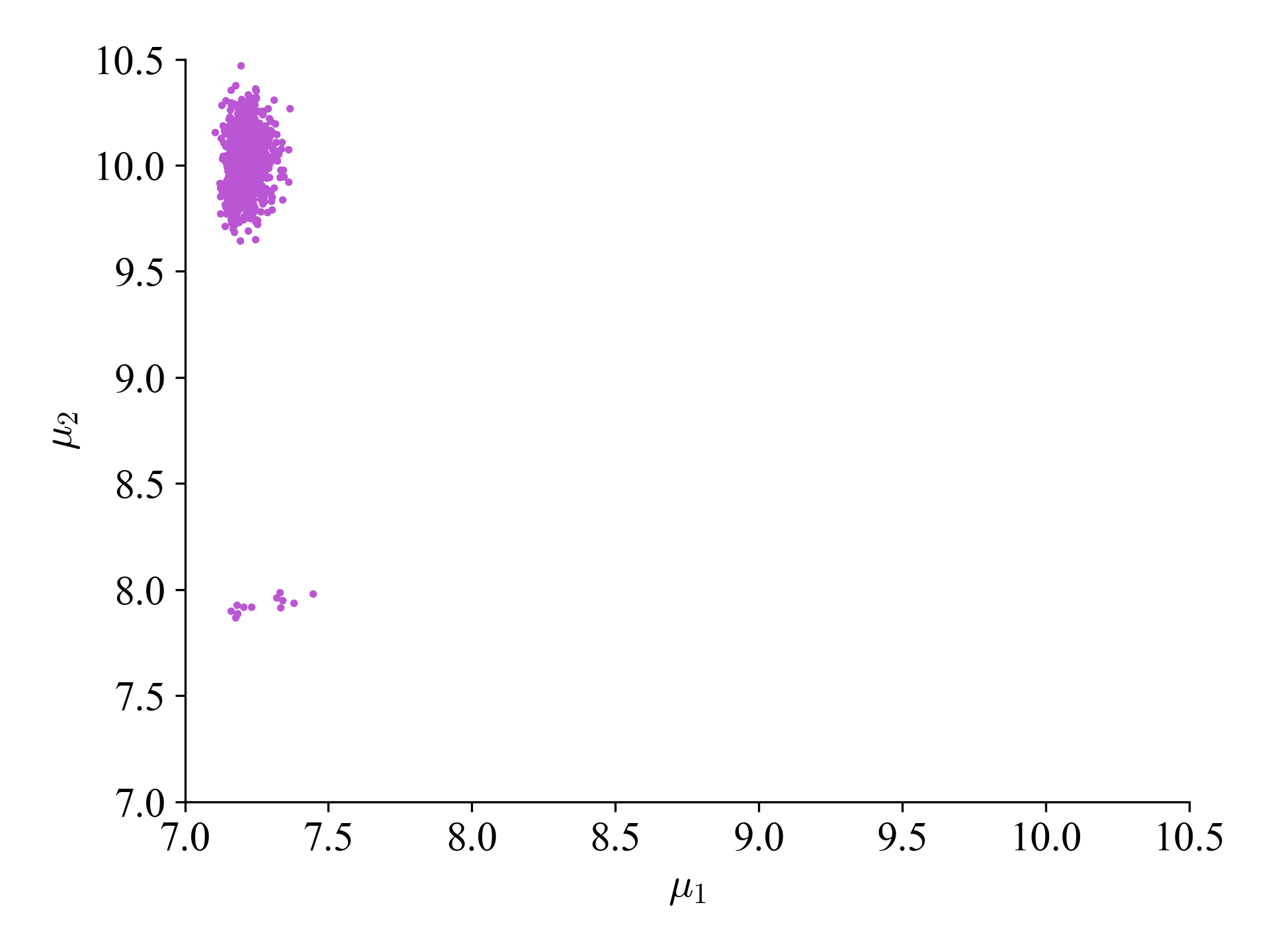}
        \caption{SMC posterior samples for Gaussian mixture model with Langevin proposal.}
        \label{fig:met_samps}
	\end{minipage}%
	\hfill
	\begin{minipage}{0.48\textwidth}
        \centering
        \includegraphics[width=0.9\textwidth]{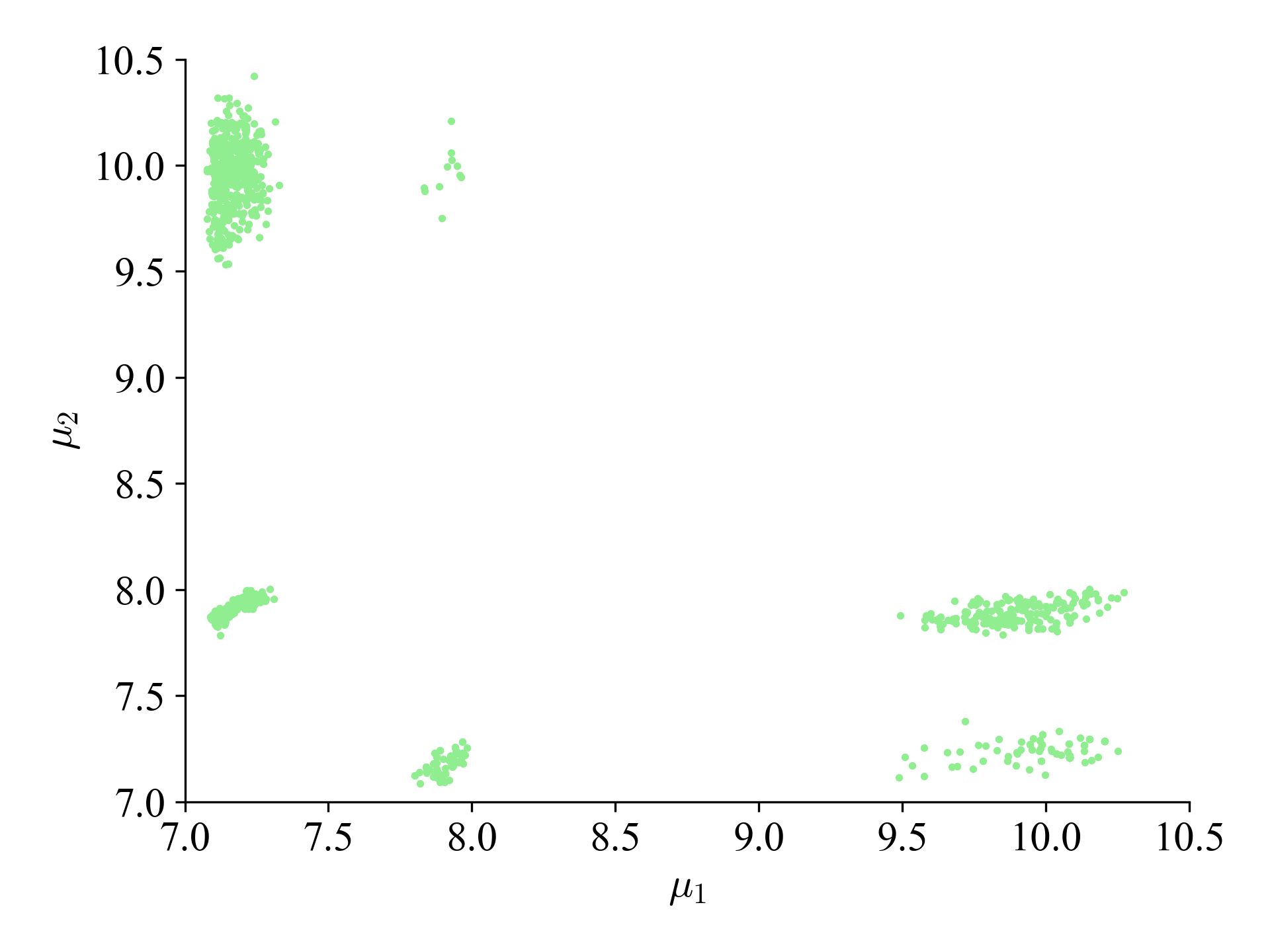}
        \caption{SMC posterior samples for Gaussian mixture model with Quasi-Newton Langevin proposal.}
        \label{fig:met_qn_samps}
	\end{minipage}
\end{figure}

In this example, the true posterior is not available analytically. Thus to compare the accuracy of particle approximations we compare the estimate of the (log) normalising constant $ \hat{Z}_T \approx Z_T =\int p(x)p(y \mid x) dx$ which is calculated sequentially $\hat{Z}_T = \hat{Z}_0 \prod_{t=1}^T \hat{Z}_\sss{t|0:t-1}$ (as described in \Cref{alg:met_smc}). The final estimate of $\log \hat{Z}_T$ for the two sequential Monte Carlo algorithms is displayed in \Cref{fig:gmm_logz} alongside the ``true'' log normalising constant calculated from an extended run of SMC with classical Langevin proposals and a very large $10000$ particles. We observe that the quasi-Newton approach is more precise and accurate. In addition, we display the posterior samples generated in the $\mu_1$ and $\mu_2$ dimensions for Langevin proposals in \Cref{fig:met_samps} and quasi-Newton Langevin proposals in \Cref{fig:met_qn_samps}. We clearly see that the use of the local Hessian approximation has allowed the quasi-Newton algorithm to explore all 6 modes whereas the classical Langevin proposals have all but collapsed to a single mode.

\section{Discussion}\label{sec:discu}

In this article, we have derived a sequential Monte Carlo technique for sampling in static Bayesian inference problems where gradient evaluations are used to form a Hessian approximation and efficiently precondition the dynamics. The Hessian approximation uses a variant of the L-BFGS algorithm that provides access to both the Hessian and inverse Hessian in a factorised square-root form \eqref{bfgs}. The approximation is cheap to compute at a cost of $O(m^2d)$, where $m$ is a tunable memory parameter that is typically set in the range 10-50. Importantly, the Hessian approximation requires no additional posterior or gradient evaluations. To our knowledge, this work provides the first application of Hessian approximations within sequential Monte Carlo methods for static Bayesian inference problems.
\par
The Hessian approximation utilises the previous $m$ states of the particle's trajectory and therefore breaks the validity of the sequential importance weights. We accept this bias in the belief that for practical problems this bias is likely to be dominated by standard Monte Carlo variance and that for suitably large $m$ and small stepsize the (projected) Hessian approximation will be increasingly accurate. In the case that the approximation is exact, we become asymptotically unbiased again.
\par
We have demonstrated the benefits of the local preconditioner in both a high-dimensional example with difficult scaling as well as a hierarchical, multi-modal example with real data.

\par
With regards to future work, a possible extension would be to investigate the potential of using a Hessian approximation within sequential Monte Carlo schemes that do not make use of an accept-reject step and therefore use a transition kernel that is not $\pi_{t-1}$-invariant. That is sequential importance weights of the form
\begin{equation*}
    w_t = w_{t-1} \frac{\pi_t(x_t) \pi_{t-1|t}(x_{t-1} \mid x_t)}{\pi_{t-1}(x_{t-1}) q_{t|t-1}(x_t \mid x_{t-1})},
\end{equation*}
where the forward kernel $q_{t|t-1}$ no longer utilises an accept-reject step and $\pi_{t-1|t}$ is an alternative backward kernel. An optimal forward kernel $q_{t|t-1}$ would directly convert a sample from $\pi_{t-1}$ into one from $\pi_t$, i.e.
\begin{equation*}
    \int \pi_{t-1}(x_{t-1}) q_{t|t-1}(x_t \mid x_{t-1}) dx_{t-1} = \pi_t(x_t).
\end{equation*}
One could investigate the possibility of combining Taylor expansions on the tempered potential \cite{Titsias2016} with a Hessian approximation to derive an approximation to an optimal forward kernel (and backward kernel \cite{DelMoral2006}).
\par
It would be desirable to implement Markovian kernels based on underdamped Langevin dynamics (or Hamiltonian Monte Carlo) as in e.g. \cite{Daviet2018, Buchholz2021}. However, there are some obstacles to the implementation of these kernels in the preconditioned case. Omission of the matrix derivatives \cite{Ma2015} in preconditioned underdamped Langevin dynamics cannot easily recover a $\pi_{t-1}$-invariant kernel even with an accept-reject step. In this work, we have therefore left the generalisation to (preconditioned) underdamped Langevin dynamics to future work. In particular, it may be possible to carefully Metropolis the dynamics with a sophisticated discretisation scheme (which goes beyond the classical leapfrog integrator followed by accept-reject method which requires a volume conservation property in the dynamics which is lost by the omission of the matrix derivatives) \cite{Girolami2011, Leimkuhler2018}.
\par
Another choice of intermediate distributions for sequential Monte Carlo is the so-called batch intermediates \cite{Chopin2012}
\begin{equation*}
    \pi_t(x_t) = p(x_t \mid y_1, \dots, y_t),
\end{equation*}
however batched intermediates do not necessarily result in smooth transitions between distributions (i.e. the $\chi^2$ distribution between consecutive distributions could be large). It is possible to combine likelihood tempering and data batching
\begin{equation*}
    \pi_t(x_t) \propto p(x_t \mid y_1, \dots, y_{r}) p(y_{r+1} \mid x_t)^{\lambda_t}.
\end{equation*}
We can now enforce smooth transitions between targets using the same effective sample size based adaptive tempering from \Cref{adapt_temp}, although we still require a routine describing how to batch the data. An interesting extension of this approach is that the aforementioned sequential Monte Carlo algorithms can be applied in online settings, where the data is received sequentially but the unknown parameters are still assumed to be static. This represents an advantage over Markov Chain Monte Carlo approaches which cannot be updated in light of further observations.

\bibliographystyle{plain}
\bibliography{references}



\end{document}